\documentclass[prl,twocolumn,aps,epsf,showpacs,superscriptaddress,longbibliography]{revtex4-2}

\usepackage[dvipsnames]{xcolor}
\usepackage{physics}
\usepackage{amsmath}
\usepackage{amsfonts}

\usepackage{graphicx}
\begin{document}

\title{Transitions in the learnability of global charges from local measurements}

\author{Fergus Barratt}
\affiliation{Department of Physics, University of Massachusetts, Amherst, MA 01003, USA}
\author{Utkarsh Agrawal}
\affiliation{Department of Physics, University of Massachusetts, Amherst, MA 01003, USA}
\author{Andrew C. Potter}
\affiliation{Department of Physics and Astronomy, and Quantum Matter Institute,
University of British Columbia, Vancouver, BC, Canada V6T 1Z1}
\author{Sarang Gopalakrishnan}
\affiliation{Department of Physics, The Pennsylvania State University, University Park, PA 16802, USA}
\author{Romain Vasseur}
\affiliation{Department of Physics, University of Massachusetts, Amherst, MA 01003, USA}

\begin{abstract}

We consider monitored quantum systems with a global conserved charge, and ask how efficiently an observer (``eavesdropper'') can learn the global charge of such systems from local projective measurements. We find  phase transitions as a function of the measurement rate,  depending on how much information about the quantum dynamics the eavesdropper has access to. For random unitary circuits with $U(1)$ symmetry, we present an optimal classical classifier to reconstruct the global charge from local measurement outcomes only. We demonstrate  the existence of phase transitions in the performance of this classifier in the thermodynamic limit.  We also study numerically improved classifiers by including some knowledge about the unitary gates pattern.

%
%
\end{abstract}
\maketitle
{\bf Introduction.} A recent breakthrough in our understanding of open quantum systems has been the discovery of measurement-induced phase transitions (MIPTs) in monitored quantum systems \cite{skinner2019measurement,liQuantumZeno2018, chanUnitaryprojectiveEntanglement2019, potterEntanglementDynamics2021}. MIPTs have been best characterized in random quantum circuits, but seem to be a generic consequence of the competition between chaotic dynamics and measurements~\cite{skinner2019measurement,liQuantumZeno2018, chanUnitaryprojectiveEntanglement2019, potterEntanglementDynamics2021,Xiangyu19,Schomerus19,gullansScalableProbes2020,jianMeasurementinducedCriticality2020,ChoiTheory20,choiQuantumError2020,gullansDynamicalPurification2020,zabaloCriticalProperties2020,Buchler20,Vijay2020Volume,LuntMBL20,Qicheng20,Turkeshi2d20,Yohei20,Szyniszewski2020,Iaconis20,yoshidaDecodingEntanglement2021,Lavasani21,turkeshiMeasurementinducedEntanglement2021, sangEntanglementNegativity2021, liRobustDecoding2021,Hsieh21,Li2021Conformal, liEntanglementDomain2021,gullansQuantumCoding2021,ippolitiEntanglementPhase2021,fanSelforganizedError2021, BAO21,DiehlPRL21,Lunt21,LavasaniPRL21,noel2021observation,Hafezi21,Nahumtrees21,Swingle21Brownian,jian2021phase,agrawalEntanglementChargeSharpening2021, barrattFieldTheory2021,liStatisticalMechanics2021a,Melko21, weinsteinMeasurementinducedPower2022,Sierant2022dissipativefloquet,Yiqiu22, baoFiniteTime2022,zabaloOperatorScaling2022,sahu2022entanglement,SierantTurkeshi22,koh2022experimental,dehghani2022neuralnetwork,zabalo2022infiniterandomness}. The best-studied MIPT, in random circuits, is a transition in the properties of a quantum state \emph{conditional} on a set of measurement outcomes. It has multiple equivalent formulations, of which the most relevant one for our purposes is as follows~\cite{gullansDynamicalPurification2020}. When the measurement rate is high, local measurements can rapidly distinguish different initial states; in this ``pure'' phase, conditional on the outcomes, an initially mixed state quickly becomes pure. When the measurement rate is low, scrambling dominates, so initially distinct states become indistinguishable by local measurements. In this ``mixed'' phase, an initially mixed state remains mixed for times that scale exponentially with system size~\cite{gullansDynamicalPurification2020}. Mixed-phase dynamics forms a quantum error correcting code in the sense that it protects initial-state information from local observers~\cite{gullansDynamicalPurification2020,choiQuantumError2020,dehghani2022neuralnetwork}. 
Studying the MIPT as formulated above requires repeated generation of the same set of measurement outcomes, which in turn requires running each circuit a number of times that grows exponentially with system size and evolution time. Experimental studies of the MIPT have therefore been limited to very small systems~\cite{noel2021observation,koh2022experimental}. 

In principle, the measurement outcomes in the pure phase suffice to distinguish any two initial states. Thus one would have a way around postselection if one could initialize the system in a mixed state, run the circuit once while recording the measurement outcomes, and use the outcomes to \emph{predict} some property of the resulting pure state that can be measured in a single shot. In the original random-circuit setting, this task is impractical, at least on a classical computer: to distinguish the two initial states, one would need to time evolve both with the specified measurement outcomes, and this is exponentially hard even with full knowledge of the unitary evolution operator and the measurement locations. (Similar challenges arise in the problem of reconstructing information from evaporating black holes~\cite{HaydenPreskill,HarlowHayden,yoshida2017efficient}.) Without such knowledge, predicting any local property of the final state is impossible: the space of possible unitaries involves arbitrary single-site rotations, so the knowledge gleaned from previous measurements is in a basis that is effectively hidden from the predictor. 

\begin{figure}
    \includegraphics[width=\linewidth]{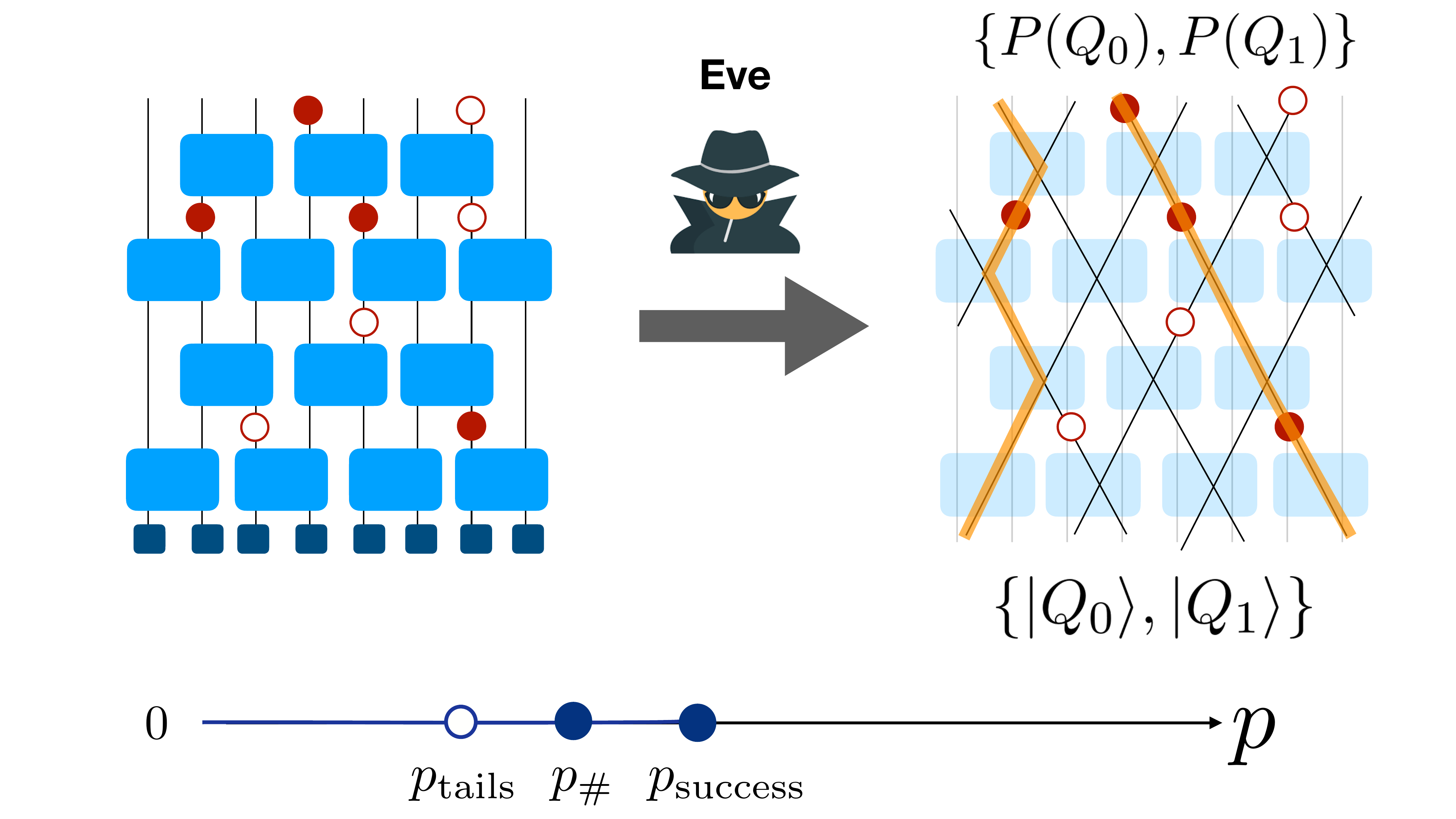}
    \caption{{\bf Setup.} An eavesdropper (Eve) attempts to reconstruct the global charge of a random quantum circuit from local charge measurements using a classical classifier. Eve can make exact predictions with success probability tending to 1 in the thermodynamic limit above the success transition $p_{\rm success}$. The success transition of the classifier is lower bounded by the charge sharpening transition $p_\#$ of the system. At $p=p_{\rm tails}$, the distribution of the probability of correct label changes shape (see text).
    \label{FigSetup} }
\end{figure}

Here, we show that constraining the unitary dynamics to have a single conserved charge (and measuring the local charge density) makes it possible to accurately predict an observable (namely the total charge) on a single run of the circuit, even without knowledge of the gates. 
We consider a one-dimensional system of $L$ qubits with a conserved $U(1)$ charge   $ Q = \sum_i q_i$, where $q_i = \left(Z_i + 1\right)/2$. We initialize the system in one of two charge states $\ket{Q_0}$, or $\ket{Q_1}$. 
We then evolve the system in time with a brickwork of random unitaries, with each time step corresponding to two layers of gates acting on even and odd sites. The gates are chosen to conserve the $U(1)$ charge, but are otherwise Haar-random~\cite{Khemani2018,Rakovszky18}. At each timestep, we allow an eavesdropper (``Eve'') to measure the local charge $q_i$ on each site of the system with independent probability $p$. At some time $t_f$ that unless otherwise specified we will take to be $t_f=L$, Eve uses the measurement record and a decoding algorithm to produce a guess of the charge of the initial state (Fig.~\ref{FigSetup}). 
Eve then shares her prediction and is told if it is correct.
Symmetric monitored quantum circuits exhibit a charge-sharpening transition at $p=p_\#$, within the mixed phase, that separates a phase where the final state conditional on the circuit and measurement outcomes has a definite charge from one where it does not~\cite{agrawalEntanglementChargeSharpening2021, barrattFieldTheory2021}. In the case where Eve has unlimited resources, she can accurately predict the outcome if $p$ exceeds $p_\#$.

Here, we argue that if Eve only has access to the measurement records, and has no knowledge about the unitary gates that were applied in each circuit except their distribution  (``eavesdropping'' scenario), the optimal decoding algorithm can be constructed by counting charge configurations consistent with the measurement outcomes the observer receives. This involves evaluating the partition sum of a classical statistical mechanics model, a task that can be efficiently performed on a classical computer.
 We also show that knowledge of the dynamics in between measurements can be used to improve the classifier (``learning'' scenario), and discuss various transitions associated with this learning problem.

{\bf Independent measurements estimate.} For small measurement probability $p \ll1$, it is natural to assume that the measurement outcomes $ \{ \vec{m} \}$ are independent. To estimate the charge, Eve can then simply use the averaged charge ${Q}_{\rm estimate} = \frac{L}{M} \sum_{n = 1}^M m_n$ with $M \sim 2 p t_f L$   the number of measurements,  and determine whether it is closer to $Q_0$ or $Q_1$. Assuming independent measurements and using the central limit theorem, the probability of success (``accuracy'') of Eve to distinguish two charges $Q_0=L/2$ and $Q_1=L/2-1$ is  
\begin{equation}
\alpha_{\rm lower \ bound}=
 \frac{1}{2} \left( 1 + {\rm erf} \sqrt{ \frac{p t_f}{L}} \right). 
\end{equation}
In general, measurement outcomes are correlated in interesting ways that can be used to improve the charge estimate, and this uncorrelated result lower bounds the accuracy of other more effective classifiers. 

{\bf Optimal classifier.} Charge conservation and locality induce correlations between measurement outcomes; accounting for these correlations allows us to outperform the independent-measurements estimate. For example, measuring three out of four legs of a gate determines the charge at the fourth, or measuring a charge $q=1$ in one of the incoming legs and $q=0$ in one of the outgoing legs fully determines the charges of the other two legs even if they are not measured. 


The constraints from charge conservation can be turned into an efficient classifier. Intuitively, in the absence of information about the underlying physical dynamics, the best Eve can do is count charge configurations compatible with the measurement outcomes, assuming that charges perform random walks with the same diffusion constant as the quantum model. 
The combinatorics of such random walks is governed by the partition function of a classical statistical mechanics model describing a Symmetric Exclusion Process (SEP)~\cite{SPITZER1970246} subject to quenched constraints from measurements, which Eve could simulate efficiently on a classical computer.
In the supplemental material~\cite{supp}, we show that this is indeed the optimal scheme, in the sense that it minimizes the misclassification probability. Formally this statistical mechanics description emerges from averaging the Born probabilities of the quantum models over the Haar unitary gates. Remarkably, the same model also emerged in the context of measurement-induced charge-sharpening phase transitions in the limit of large onsite Hilbert space dimension~\cite{agrawalEntanglementChargeSharpening2021,barrattFieldTheory2021}.

To represent the possible charge dynamics of Haar-random circuits, consider the time evolution of the (classical) probability distribution over computational basis states. 
The initial distribution over basis states in the Haar model is uniform over all states of a fixed charge $Q$. 
Represent this initial probability distribution of charge possibilities by a vector of size $2^L$: $  |\phi(0)) = |Q) = \binom{N}{Q}^{-1} \sum_{i: Q(i) = Q} |i)$, where we use the ket-like notation $|\psi)$ to denote a probability vector.
At each time interval, each charge can either hop or remain at the same position.
Averaged over the Haar distribution for the unitaries, each possibility has probability $1/2$. 
The update to the probability distribution $|\phi(t))$ represented by the unitary at position $i$ can be represented by the application of the transfer matrix of the symmetric exclusion process (SEP),
\begin{equation}
    T_i= \begin{pmatrix}
        1 & 0 & 0 & 0 \\ 
        0 & 1/2 & 1/2 & 0 \\ 
        0 & 1/2 & 1/2 & 0 \\
        0 & 0 & 0 & 1
    \end{pmatrix},
\end{equation}
to the probability distribution $|\phi(t))$. 

Every time the quantum state is measured at a site $k$, the corresponding probability vector $|\phi(t))$ must be modified such that all states inconsistent with the measurement outcome on that site have probability $0$. 
This can be achieved by applying the projector onto the correct measurement outcome. 
In applying the projector, the $1$-norm of the probability distribution decreases, by an amount corresponding to the fraction of trajectories that were inconsistent with that measurement outcome.

\begin{figure*}[t]
    \centering
    \includegraphics[width=\linewidth]{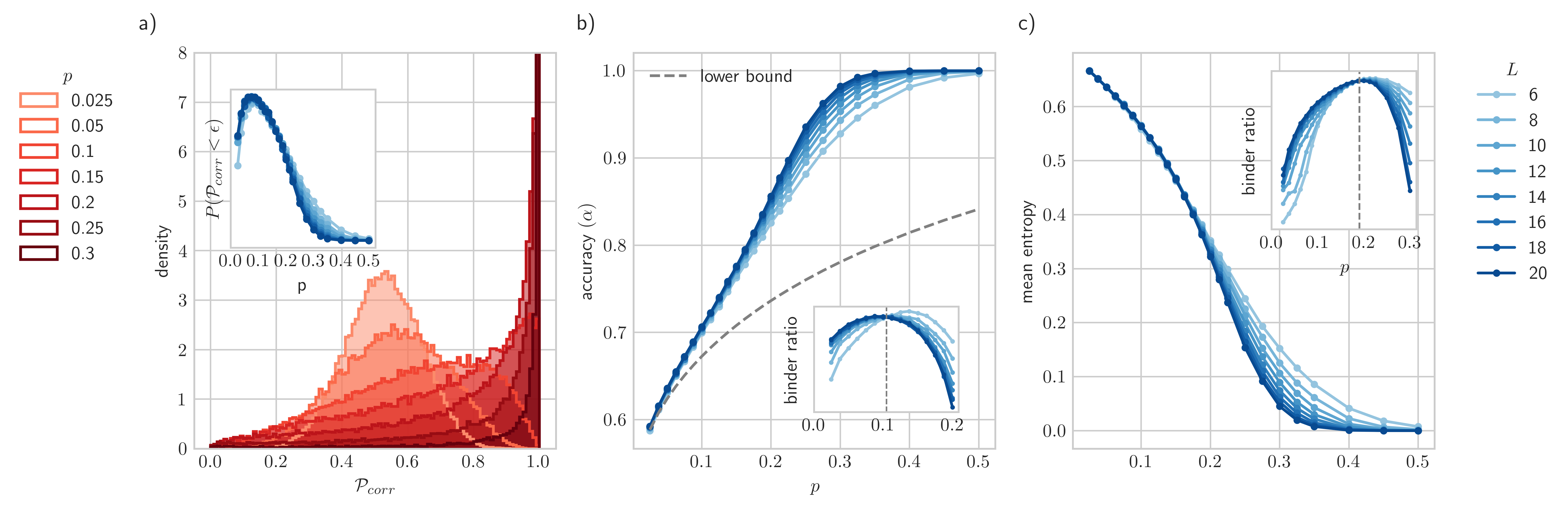}
    \caption{{\bf Optimal classifier}. a) Probability distribution of the `probability associated with the correct charge label', ${\mathcal P}_{\mathrm{corr}}$. Inset: Weight in lower tail of distribution of ${\mathcal P}_{\mathrm{corr}}$, $\epsilon=0.4$. The distribution changes through three distinct regimes:  approximately Gaussian for  $0 < p < p_{\rm tails} \simeq 0.1$, power-law for $p =p_{\rm tails} $, exponential for $p>p_{\rm tails}$. b) Accuracy of the classifier. Inset: The Binder ratio shows a crossing at $p_{\rm tails} \simeq 0.1$. c) The mean entropy as an order parameter for the success transition, above which Eve can systematically make  accurate predictions in the thermodynamic limit. Inset: the Binder ratio has a crossing at $p_{\rm success}\simeq 0.2$.
    }\label{fig:pcorr} 
\end{figure*}

Define $T(\vec{m}_t)$ 
to be the linear operator that updates the probability vector from time $t$ to $t+1$, given the constraints represented by the measurements. 
After time $t_f$, we have a state $|\phi(t_f)) = \prod_{t=1}^{t_f} T(\vec{m}_t) |Q)$ representing the uniform distribution over all charge trajectories that are consistent with both the measurement outcomes and the charge $Q$. 
The $1$-norm of the state represents the fraction of all possible trajectories of the charges in the Haar-random circuit that are compatible with the constraints -- and can be found by the dot product of the probability vector with the (unnormalised) uniform distribution over \emph{all} states $|1) = \sum_i |i)$,
\begin{equation}
    P(\{ \vec{m} \}| Q) =  (1| \prod_{t=1}^{t_f} T(\vec{m}_t)  |Q). \label{eq:dot_product}
\end{equation}
Eve can then use $P(Q | \{ \vec{m} \}) = P(\{\vec{m}\} | Q) / (P(\{\vec{m}\} | Q_0) + P(\{\vec{m}\} | Q_1))$ to determine which charge is more likely given a set of measurement outcomes  $\{ \vec{m} \}$.

{\bf Efficiency.} This probability can be found for a given circuit realisation by explicitly evolving the state $|Q)$ using a full representation of the probability vector. Naively, this algorithm scales as $\mathcal{O}(\mathrm{poly}(2^L))$. We can do better by noticing that the circuits and measurements (since they are not determined by properties of the time evolving state, but by the separate dynamics in the Haar circuit) represent a set of predetermined linear operations applied to the initial state. 
Instead of applying them to the (highly entangled) state $|Q)$ , we can apply them in reverse to the (weakly entangled) state $|1)$ as $(\psi(t_f)| = (1| \prod_{t=1}^{t_f} T(\vec{m}_t)$.
Because of the non-unitary of the SEP dynamics, the entanglement growth generated by the transfer matrix is significantly lower than that in the Haar circuit, and so the system can be represented by a Matrix Product State (MPS) with a bond dimension that grows sublinearly in time. 
This allows simulation of systems up to large sizes using MPS algorithms like TEBD~\cite{vidalEfficientClassical2003,SCHOLLWOCK201196}.

The state $|Q)$ cannot be efficiently represented on a classical computer, and so we cannot efficiently compute the dot-product in Eq.~\ref{eq:dot_product}.
We can however, efficiently sample from $( \psi(t_f)|$ (since it has a low bond dimension MPS representation) to produce an estimate of $P(Q)$. We also note that this statistical mechanics problem has positive Boltzmann weights, and could be simulated efficiently using Monte Carlo methods. 

{\bf Success transition.} In order to probe the performance of the classifier, we consider its performance on $N=40,000$ random Haar measurement records.
Half the records are generated from initial state $\ket{Q_0} = \ket{L/2}$, i.e. the uniform distribution over bitstrings at half filling, the other half from $\ket{Q_1} = \ket{L/2-1}$.

The task assigned the classifier is determining which state the record was generated from. 
Given the probabilities $P(Q_1|\{\vec{m}\}), P(Q_0|\{\vec{m}\})$ from the stat.~mech.~model, the classifier chooses the $Q$ such that $P(Q|\{\vec{m}\})$ is maximal.
The accuracy $\alpha$ of this classifier as a function of measurement rate and system size is presented in Fig.~\ref{fig:pcorr}b).
The classifier gets better at solving the task as the measurement probability increases, as expected. 

To get a better sense of the distribution of the classifier predictions across different measurement outcomes, we define the `probability of correct label' ${\mathcal P}_{\mathrm{corr}}$ as follows. Suppose the initial state has charge $Q^*$ (unknown to Eve). Eve begins with no information about the charge, then updates her probabilities based on the observed measurement outcomes. Her posterior probability for the correct charge label $Q^*$ is denoted ${\mathcal P}_{\mathrm{corr}}$: i.e., ${\mathcal P}_{\mathrm{corr}} = P(Q^*|\{ \vec{m} \})$. 
Note that since Eve is told the value of $Q^*$ at the end of each run, ${\mathcal P}_{\mathrm{corr}}$ is measurable for each run, so
Eve has access to the entire probability distribution $P({\mathcal P}_{\mathrm{corr}})$, plotted in Fig.~\ref{fig:pcorr}a). 
In terms of $\mathcal{P}_{\rm corr}$, the accuracy (Fig.~\ref{fig:pcorr}b) is $P({\mathcal P}_{\mathrm{corr}} > 0.5)$. 

The entropy of the binary distribution $\lbrace {\mathcal P}_{\mathrm{corr}}, 1- {\mathcal P}_{\mathrm{corr}} \rbrace $ corresponds to the confidence of the classifier in its decision -- irrespective of the ground truth label. The binder ratio of the entropy has a crossing at $p_{\rm success} \approx 0.2$, which corresponds to a ``success transition''. Above $p_{\rm success}$, Eve can reconstruct the charge of the system {\rm exactly} (i.e. with success probability tending to $1$ as $L\rightarrow \infty$). Interestingly, this success transition in the classifier also has an interpretation as a charge-sharpening transition in a charge-conserving model with large on-site Hilbert space~\cite{ agrawalEntanglementChargeSharpening2021}, and its critical properties are Kosterlitz-Thouless-like~\cite{barrattFieldTheory2021}.  
In general, we have the constraint $p_\# \leq p_{\rm success}$, with $p_\# \simeq 0.09$ for qubit systems~\cite{agrawalEntanglementChargeSharpening2021}: classifiers can only make systematic, exact predictions above the sharpening transition.

The full distribution of ${\mathcal P}_{\mathrm{corr}}$  (Fig.~\ref{fig:pcorr}a)) reveals a richer structure.  For low $p<p_{\rm tails} \simeq 0.1 $, the measurement rate is insufficient for the observer to fix the charge, resulting is an approximately Gaussian distribution of ${\mathcal P}_{\mathrm{corr}}$. Around $p=p_{\rm tails} $, the tails of distribution of ${\mathcal P}_{\mathrm{corr}}$ empirically change from Gaussian to power-law like. This apparent transition in the tails of the distribution can also be detected from the Binder ratio of ${\mathcal P}_{\mathrm{corr}}$ (inset of Fig.~\ref{fig:pcorr}b)), and from the power-law shape of the distribution of ${\mathcal P}_{\mathrm{corr}}$, see inset of Fig.~\ref{fig:pcorr}a). Note that even though the quantity $1 - \mathbb{E} ({\mathcal P}_{\mathrm{corr}})$ (where $\mathbb{E}(\ldots)$ denotes an average) is itself an order parameter for the success transition~\cite{supp}, the heavy-tailed distribution allows its Binder ratio to cross at a measurement rate that is different from $p_{\mathrm{success}}$. 
%
%
It would be interesting to analyze these tail transitions further in future work. 



\begin{figure}[t] 
    \includegraphics[width=\linewidth]{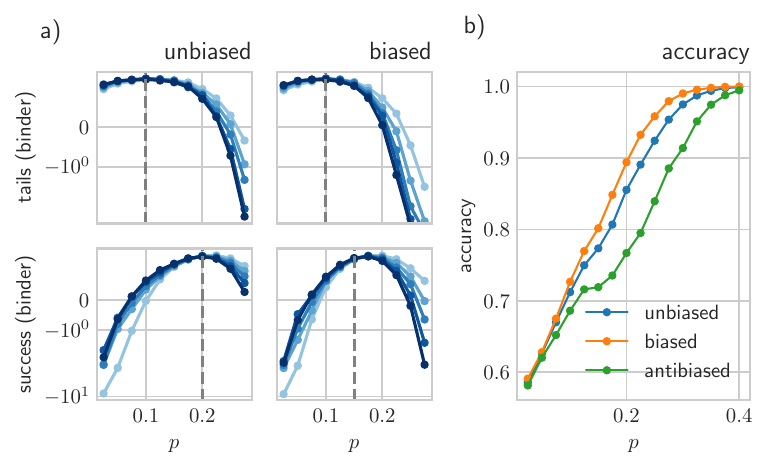}
    \caption{{\bf Biased classifier}. (a)  The success transition is lowered by including information about the hopping probabilities in the classifier (biased model), while the accuracy transition remains unchanged (b) The classifier accuracy $\alpha$ is improved by including information about the hopping probabilities in the classifier. \label{FigBiased}}
\end{figure}


{\bf Biased classifier.} While the above classifier is optimal without additional knowledge about the circuit,  it can be improved if Eve has some information about the underlying dynamics of the system (learning scenario). Let us assume now that for each run of the experiment, Eve receives the set of measurement outcomes and locations $\{\vec{m}, \vec{x}\}$, \emph{and} the details of the unitary gates $\{U_{it}\}_{\forall i, t}$that were applied to generate this measurement record. 

There is a trivial, optimal, exponentially classically hard algorithm -- the observer can run the circuit starting from $\ket{Q_0}$, and $\ket{Q_1}$, measure the charge in the locations specified and count how many times the measurement record $\vec{m}$ arises. We expect this algorithm to succeed above the charge-sharpening transition ($p>p_\#$). A more interesting task is to find an \emph{efficient} classical algorithm that improves on the zero-knowledge classifier above.
Define the hopping amplitude of a unitary $h(U) = \abs{\bra{01}U\ket{10}}^2$.
This has the  properties $\overline{h(U)} = \frac{1}{2}$, where the overline indicates average over Haar,  $h({\rm SWAP}) = 1$ and $h(\mathbb{I}) = 0$.
We can then modify the classifier above using the disordered hopping probabilities:
\begin{equation}
    T_i(t) = \begin{pmatrix}
        1 & 0 & 0 & 0 \\ 
        0 & p_{it} & 1-p_{it} & 0 \\
        0 & 1-p_{it} & p_{it} & 0 \\
        0 & 0 & 0 & 1 \\ 
    \end{pmatrix}
\end{equation}
and three classifiers -- \emph{unbiased}, with $p_{it} = \frac{1}{2}$, \emph{biased}, with $p_{it} = 1-h(U_{it})$, and \emph{antibiased}, with $p_{it} = h(U_{it})$. 
The unbiased model is the same as before, the biased model has hopping amplitudes that match the Haar-random circuit, and the antibiased model has hopping amplitudes that are opposite to the biased one. The performance of the biased classifier are summarized in Fig.~\ref{FigBiased}. As expected, the biased model improves the accuracy of the classifier, although it does not change the location of the accuracy transition. The biased model has a lower success transition at $p_{\rm success} \simeq 0.15$, closer to the fundamental sharpening bound $p_\# \simeq 0.09$ . 
Additional results on the antibiased classifier are presented in the supplemental material~\cite{supp}.

{\bf Discussion.}
When the measurement rate $p$ is high enough, the history of measurement outcomes suffices to distinguish any two initial states, and quantum information in the system is unprotected from its environment.
Even if the environment contains this ``which-state'' information, extracting it and predicting the state of the system naively requires (a)~full knowledge of the circuit, and (b)~${\rm e}^{L}$ resources for a chain of length $L$. 
We showed here that, for local dynamics with a conservation law, one can extract which-state information with polynomial overhead and with no knowledge of the gates in the circuit, by exploiting hydrodynamic correlations between measurements at different times.
The threshold for in-practice extractability, $p_{\rm success}$, exceeds that for in-principle extractability, $p_{\#}$. An interesting open question is whether, between these thresholds, the charge can be extracted given full knowledge of the circuit but only polynomial resources on a classical computer. 
%

Our setup is analogous to the problem in black-hole physics where Alice drops a qubit into an old black hole and Bob attempts to reconstruct it from the emitted radiation~\cite{HaydenPreskill}. 
The question addressed here is distinct (but in a sense ``dual'') to the problem of finding optimal decoders in the volume-law phase of the standard MIPT~\cite{gullansDynamicalPurification2020,choiQuantumError2020,dehghani2022neuralnetwork}. There, the measurement record contains no information about the encoded qubit, but is instead used to find a unitary operation \emph{on the circuit} that unscrambles the input qubit. 
In our setup, the input qubit has leaked into the environment, and the task is instead to unscramble the environment. It would be interesting to extend our results to this decoding problem in symmetric circuits, and to explore consequences for covariant error correction~\cite{faistContinuousSymmetries2020,kongNearoptimalCovariant2021}.

{\it Acknowledgements.} We thank Kun Chen, Michael Gullans, David Huse, Jed Pixley, Justin Wilson, and Aidan Zabalo for useful discussions and collaborations on related projects, and Michael Gullans for helpful comments on this manuscript.
We acknowledge support from the NSF through grants DMR-1653271 (S.G.), DMR-1653007 (A.C.P.), the Air Force Office of Scientific Research under Grant No. FA9550-21-1-0123 (F.B. and R.V.), and the Alfred P. Sloan Foundation through Sloan Research Fellowships (A.C.P. and R.V.).

\bibliography{decoding}

%
%
%
%
%
\end{document}


\title{Supplemental Material for ``Transitions in the learnability of global charges from local measurements''}
\author{Fergus Barratt}
\affiliation{Department of Physics, University of Massachusetts, Amherst, MA 01003, USA}
\author{Utkarsh Agrawal}
\affiliation{Department of Physics, University of Massachusetts, Amherst, MA 01003, USA}
\author{Andrew C. Potter}
\affiliation{Department of Physics and Astronomy, and Quantum Matter Institute,
University of British Columbia, Vancouver, BC, Canada V6T 1Z1}
\author{Sarang Gopalakrishnan}
\affiliation{Department of Physics, The Pennsylvania State University, University Park, PA 16802, USA}
\author{Romain Vasseur}
\affiliation{Department of Physics, University of Massachusetts, Amherst, MA 01003, USA}
\date{\today}
\pacs{}
\maketitle
\tableofcontents

\section{Optimality of the stat. mech. decoder}
It is a known result \cite{bishopPatternRecognition} that the optimal classifier (in the sense of minimizing the misclassification probability in the infinite data limit) for a dataset consisting of data, label pairs $(X, Y)$ drawn from joint distribution $p(x, y)$ with $x,y\in \{0,1\}$, is given by the Bayes' classifier
\begin{equation}
    h(x) = \mathrm{argmax}_y \left( p(y | x)\right).\label{eq:optimal_h}
\end{equation}
Here $h(x)$ is a classifier assigning labels $y \in \{0, 1\}$ to datapoints $x$.
Colloquially, the best we can do is to assign labels to datapoints such that they maximise the probability that that label would have been generated in the data distribution given that datapoint. 

In order to perform optimally at the classification task, we need efficient access to the distribution $p(y | x)$. 
In machine learning applications, the problem is most commonly that the data distribution $p(y| x)$ is unknown, and we must therefore model it in order to make appropriate predictions. 
In the setting considered here, the problem is different -- we know the full joint distribution of the data and the labels, and have expressions for all of the marginals. 
The task is to enumerate the settings in which knowledge of that distribution gives rise to a classically simulable classifier.

\subsection{Data distribution}
Introduce a label $y$, and define two states $\ket{Q_y}$, with $y \in \{0, 1\}$ the label of the initial state.
Given an initial state $\ket{Q_y}$, we draw a set of unitaries $u$ from independent haar distributions at each location in a brick wall pattern. 
We then perform measurements of the local charge at random space time positions.
Call this measurement record $m$ (consisting of both the locations of the measurements and their outcomes).
Then the joint distribution of $m$, $y$, $u$ can be decomposed as follows:
\begin{equation}
    p(m, y, u) = p(m | y, u) p(y|u) p(u) =  ||C(m, u)|Q_y\rangle ||^2 p(y) p(u).\label{eq:inefficient}
\end{equation}
Where $C(m, u)$ is the brick wall circuit, with the projectors corresponding to the measurement outcomes in $m$ applied at the correct positions, and we have used that $p(y | u) = p(y)$, i.e. that unitary realisations and choices of initial states are independent. 

The total dataset is sampled from $p(m, y, u)$, and consists of $(m, y, u)$ triples. 
The general goal is to investigate the conditions under which knowledge of $(m, u)$ (or incomplete knowledge thereof) suffices to determine the label $y$. 
From Eq.~\ref{eq:optimal_h} we can give the optimal classification rule for $(m, u)$ pairs
\begin{equation}
    h(x, u) = \mathrm{argmax}_y \left( p(y | x, u)\right).,\label{eq:optimal_h}
\end{equation}

In this instance, evaluating $\eta(x, u)$, which is required for assigning the optimal label (in terms of the classifier accuracy) requires evaluating Eq. ~\ref{eq:inefficient} for fixed $(x, u)$. 
Determining the probability of a specific measurement outcome $m$ for a fixed measurement outcome $u$ is exponentially hard on a classical computer since it requires full simulation of the dynamics of the quantum wavefunction in a volume law phase.
Interestingly, it is also in principle exponentially hard on a quantum computer. 
An arbitrary measurement outcome is exponentially unlikely to arise (in the number of measurements $N$, which scales in our setup as $O(L)$).
As such, exponentially many samples are required to perform the classification optimally. 
We do not investigate here the performance of non-optimal classifiers, although preliminary experiments with neural network classifiers have not shown promising results. 

\subsection{Classifying without knowledge of the unitary gates}
Despite the problems with classically performing the optimal classification, in the main paper we argue that given access to only a subset of the information contained in the unitary realisation $u$ we can perform a conditionally optimal classification (i.e, the best we can do given access to only that information). 
We first consider the case where we ignore the unitaries completely for each measurement record and try to determine the charge only from the information contained in the locations and outcomes of measurements. 
The dataset with the $u$ realisations ignored is the same as that generated by sampling from the \emph{marginal} distribution
\begin{equation}
    p(m, y) = \int \dd{u} p(m | y, u) p(y) p(u) =  \mathbb{E}_u \left[||C(m, u)|Q_y\rangle ||^2 p(y)\right]. \label{eq:marginal}
\end{equation}
Consider the case of balanced, binary classes $p(y) = 1/2, y \in \{0, 1\}$, then:
\begin{equation}
    p(y| m) = \frac{p(m,y)}{\sum_{y'}p(m,y')}= \frac{\mathbb{E}_U \left[||C(m, u) \ket{Q_y}||^2\right]}{\mathbb{E}_U \left[ ||C(m, u) \ket{Q_0}||^2+||C(m, u) \ket{Q_1}||^2 \right]}.\label{eq:optimal2}
\end{equation}
We will show in Sec.~\ref{sec:stat_mech} that $p(m, y) = \mathbb{E}_U \left[ || C(m, u) |Q_y \rangle ||^2 \right] = p_{sm}(m, y)$ is precisely the probability of the measurement record $m$ occuring in a constrained SEP model. 
From Eq.~\ref{eq:optimal2}, we can see that the optimal classification rule is therefore to pick the class label that maximises this probability for a given measurement record.
This is the classification rule that we investigate in the main paper. 

Eq.\ref{eq:optimal2} also demonstrates the the transitions outlined in the paper for the case in which we use no information about the unitaries are in some sense properties of the stat.~mech.~model alone. 
In particular, we can generate born weighted measurement records from Eq.~\ref{eq:marginal} by running the stat.~mech.~model simulation starting from the state $|Q_y )$ and sampling at each measurement location an outcome from the marginal distribution of measuremnent outcomes at that site. 
This is efficiently simulable with TEBD for low entanglement initial states \cite{barrattFieldTheory2021} (we demonstrate that the high entanglement states discussed for generality in the main paper are not important to the concepts presented in Sec.~\ref{sec:variations}). 
\subsection{Derivation of the Stat.~Mech.~model}\label{sec:stat_mech}
The probability of a measurement record-label pair $m, y$ arising in the dataset is given by (Eq.~\ref{eq:av_circ})
\begin{equation}
    p(m, y) = \int \dd{u} p(m | y, u) p(y) p(u) =  \mathbb{E}_u \left[||C(m, u)|Q_y\rangle ||^2 p(y)\right]
\end{equation}
Alternatively, using two copies of the circuit $C$:
\begin{align}
    \mathbb{E}_u \left[||C(m, u)|Q_y\rangle ||^2 p(y)\right] &= \mathbb{E}_u \left[\sum_k \bra{k} \otimes \bra{k} C(m, u) \otimes C^* (m, u) |Q_y\rangle \otimes | Q_y \rangle  p(y)\right]. \\
                                                             & =  \sum_k \bra{k} \otimes \bra{k} \mathbb{E}_u \left[C(m, u) \otimes C^* (m, u) \right] |Q_y\rangle \otimes | Q_y \rangle  p(y). \label{eq:av_circ}
\end{align}
We can see that the key object is the averaged operator
\begin{equation}
    \mathbb{E}_u \left[C(m, u) \otimes C^* (m, u) \right].
\end{equation}
Each unitary in the circuit is $U(1)$ symmetric and haar random. In terms of total charge sectors $0, 1, 2$, and the projectors thereon ($P_0, P_1, P_2$), the unitary at time $t$ and spatial index $i$ can be written: 
\begin{equation}
    U_{it} = e^{i\theta_0} P_0 + U_1 P_1 + e^{i \phi_2}P_2.
\end{equation}
At each spacetime position in Eq.~\ref{eq:av_circ}, the following operator is applied to both copies of the replicated initial state: 
\begin{equation}
    \mathcal{T}_{it} = \mathbb{E}_{U_{it}} \left[U_{it} \otimes U_{it}^* \right] = P_0 \otimes P_0 + \frac{1}{\sqrt{2}} \begin{pmatrix} 1 & 1 \\ 1 & 1 \end{pmatrix} \otimes \begin{pmatrix} 1 & 1 \\ 1 & 1 \end{pmatrix} P_1 \otimes P_1 + P_2 \otimes P_2, \label{eq:proj}
\end{equation}
where we have performed Haar averages within the different charge sectors.
Since Eq.~\ref{eq:proj} projects out any permutation non-symmetric contribution, all the dynamics of in Eq.~\ref{eq:av_circ} takes place within the symmetric subspace of the doubled Hilbert space. 
We can write the dynamics in terms of a basis for this space: 
\begin{equation}
    \Pi \ket{i} \otimes \ket{i} \rightarrow |i)
\end{equation}
where $\Pi$ is the projector onto the symmetric subspace. 
In this doubled basis, each doubled two site operator $\mathcal{T}_{it}$ is replaced by a single operator $T$ with the following components
\begin{equation}
    {T}_{jk} = \bra{j} \otimes \bra{j} \mathcal{T} \ket{k} \otimes \ket{k} = \begin{pmatrix} 1 & 0 & 0 & 0 \\ 0 & \frac{1}{2} & \frac{1}{2} & 0 \\ 0 & \frac{1}{2} & \frac{1}{2} & 0 \\ 0 & 0 & 0 & 1 \end{pmatrix}_{jk}
\end{equation}
whose form is exactly that of the transfer matrix of the symmetric exclusion process. 
Define $T(m)$ to be the circuit with all of the unitary matrices replaced by these transfer matrices, with projectors at each position and onto each outcome indicated by $m$. 
Define also $(1| = \sum \bra{k} \otimes \bra{k}$. 
Then,  Eq.~\ref{eq:av_circ} is equal to:
\begin{equation}
    p(x, y) = p(y) (1| T(x) |Q_y)
\end{equation}
This distribution function can be efficiently computed using the methods described in the main text, and from this the optimal classification scheme follows naturally.

\subsection{Bayesian Model Averaging}
Since the dataset under consideration consists of $(m, y)$ pairs, the $u$ records have been marginalised already, and the optimal classifier is as described above.
The classifier described above can be understood as combining multiple models of the data distribution (i.e., those with different values of $u$) according to the principles of Bayesian Model Averaging~\cite{hoetingBayesianModel1999, hinneConceptualIntroduction2020}.

One could consider, instead of the practise described in the above, choosing the maximum likelihood unitary realisation for each datapoint. 
The classification rule can be described as follows:
\begin{equation}
    h(x) = \mathrm{argmax}_y \left( p(y | x, u^*) \right)
\end{equation}
where $u^*$ is the most likely model given the data
\begin{equation}
    u^* = \mathrm{argmax}_u \left(p(u|m, y)\right).
\end{equation}
The proof outlined above shows that this practise cannot improve over the stat. mech. classifier outlined in the previous section, since it includes no more information about the model than that classifier. 
An alternative explanation for this difference in performance comes from the literature on Model Selection. 
Consider each value of $u$ as specifying a different model.
Given access to multiple models and their likelihood, one can do better than choosing the maximum likelihood model (according to a logarithmic scoring rule \cite{madiganModelSelection1994}) by performing Bayesian Model Averaging (BMA).
\begin{equation}
    h(x) = \mathrm{argmax}_y \left( \int \dd{u} p(y | x, u) p(u|m, y) \right)
\end{equation}
In which the posterior probabilities of each label associated with each model are combined with weights given by each models likelihood -- i.e. more likely models are taken into account with greater weight than less likely models. 
This process of weighting each of the models in the set is equivalent to our marginalised stat. mech. classifier. 

\begin{figure}[h]
    \centering
    \includegraphics{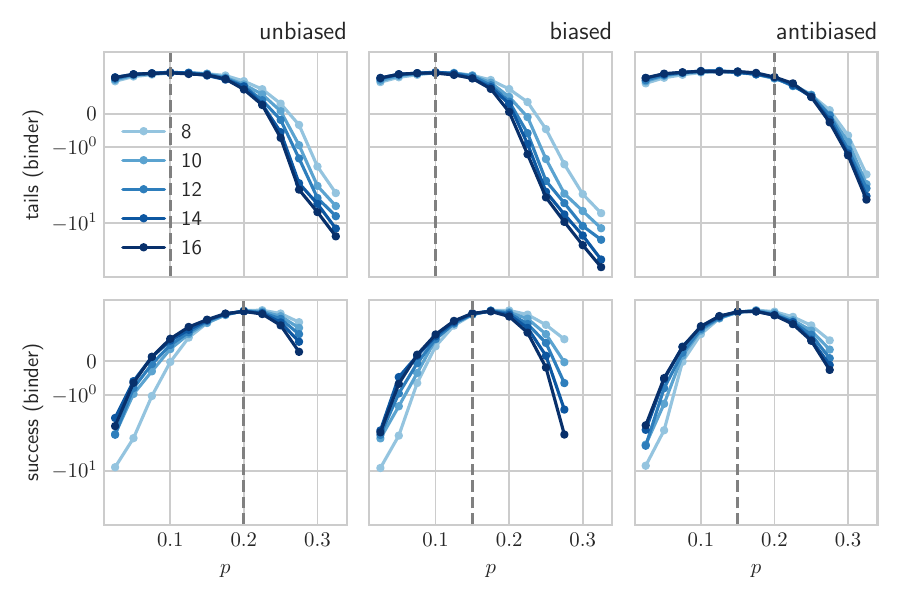}
    \caption{Transitions in the antibiased problem. Each column shows one decoder setup (unbiased, biased, antibiased) and each row shows the behaviour of the crossing in the binder ratio associated with each transition (tails, success). The data show that biasing the decoder using data from the unitary realisations improves the performance of the decoder.  }
    \label{fig:antibiased}
\end{figure}
\begin{figure}
\centering
    \includegraphics[width=0.7\linewidth]{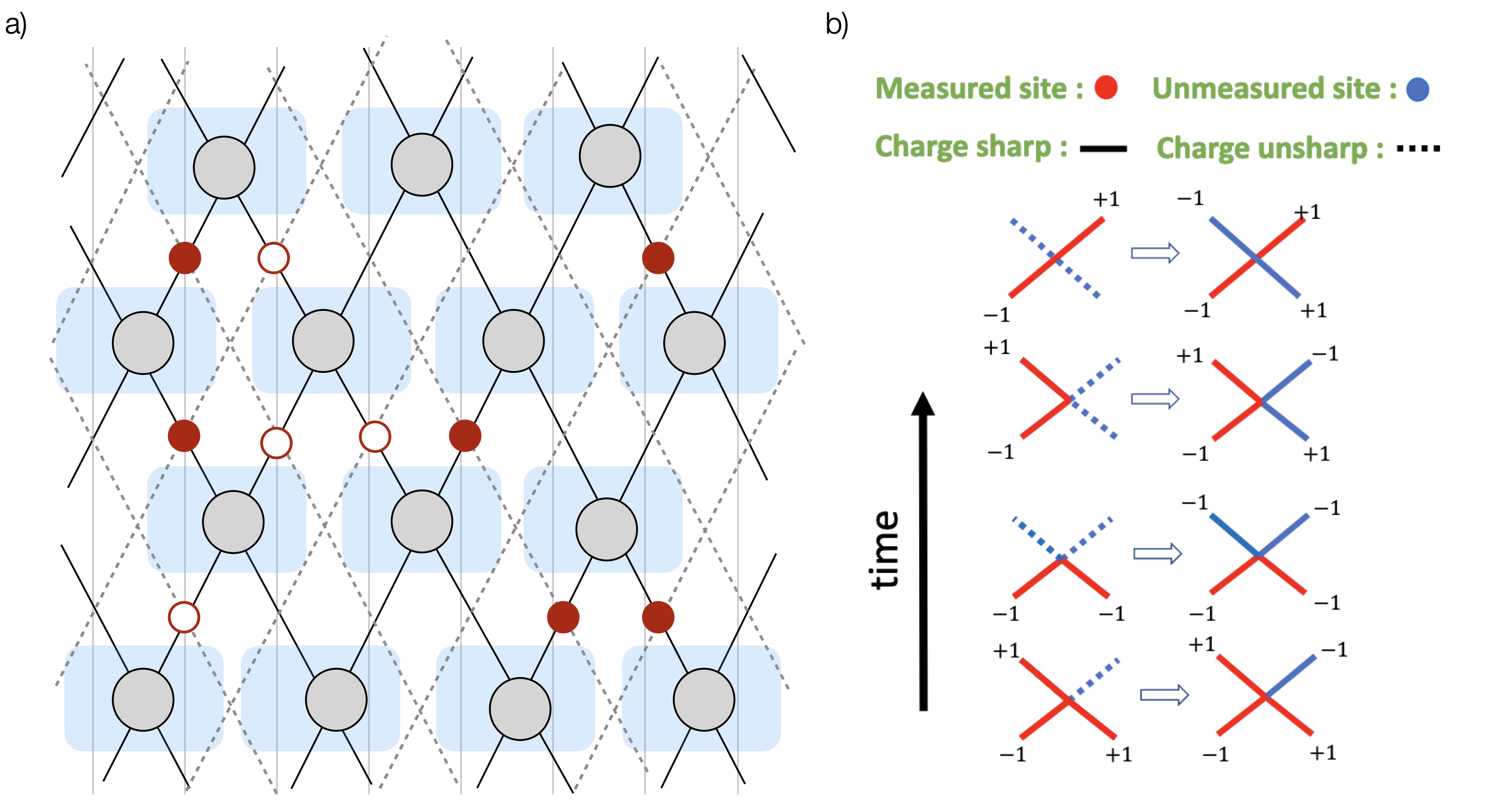}
    \caption{a) Illustration of the percolation setup. The primal lattice of unitary gates is indicated with solid black lines, whereas the dual lattice is indicated in dashed gray lines. Gates are indicated with blue squares overlaid with gray circles, measurements of $1$ with empty red circles, and of $1$ with filled red circles. Shown is a configuration in which a percolating cluster exists on the dual lattice but not on the primal lattice. b) Loops on the dual lattice (surrounding vertices on the primal lattice) allow filling out unmeasured outcomes. }
    \label{fig:dual}
\end{figure}
\begin{figure}[ht]
    \centering
    \includegraphics[width=\linewidth]{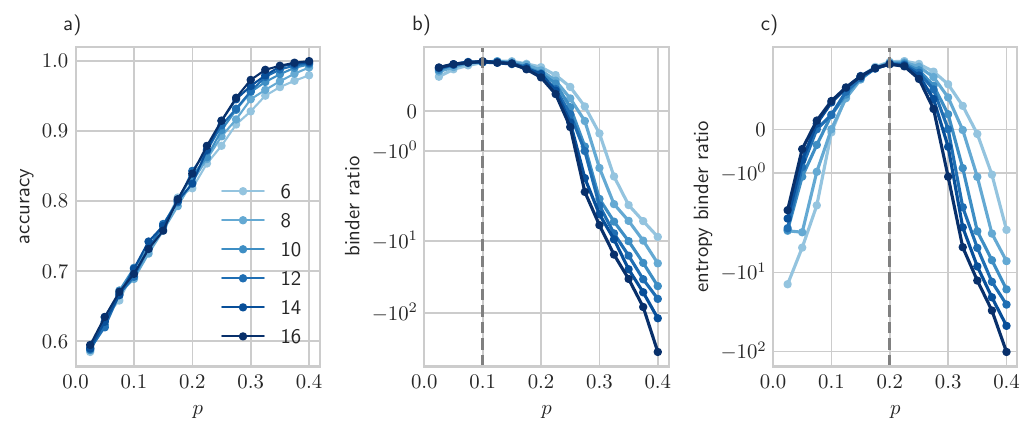}
    \caption{Transitions in detection in circuits starting from either a N\'eel state, or a N\'eel state with a single spin flip. See Fig.~1 in main text for comparison to the same quantities computed for initial states $\ket{Q_0}$ and $\ket{Q_1}$ balanced real superpositions over bitstrings with the same charge. The key observations don't depend on the specific details of the initial state - just its charge. }
    \label{fig:neel}
\end{figure}
\begin{figure}
    \centering
    \includegraphics{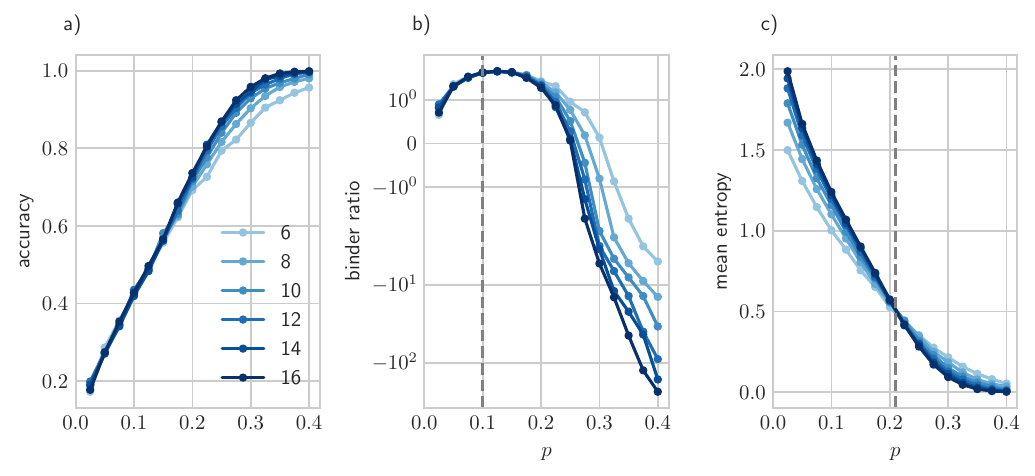}
    \caption{Detector transitions in the L class classification problem. a) Accuracy vs. $p$. The accuracy increases with increasing $p$ (and increasing $L$. b) The binder ratio of the accuracy - defined as the kurtosis of the distribution of boolean correct/incorrect values. This exhibits a crossing at $p=p_{\rm tails} \sim 0.1$. c) The mean entropy exhibits a crossing at $p_{\rm success}$ (rather than order parameter like behaviour) since its behaviour is now extensive with $L$ in the fuzzy phase.}
    \label{fig:l_class}
\end{figure}
\begin{figure}[t]
    \centering
    \includegraphics[width=\linewidth]{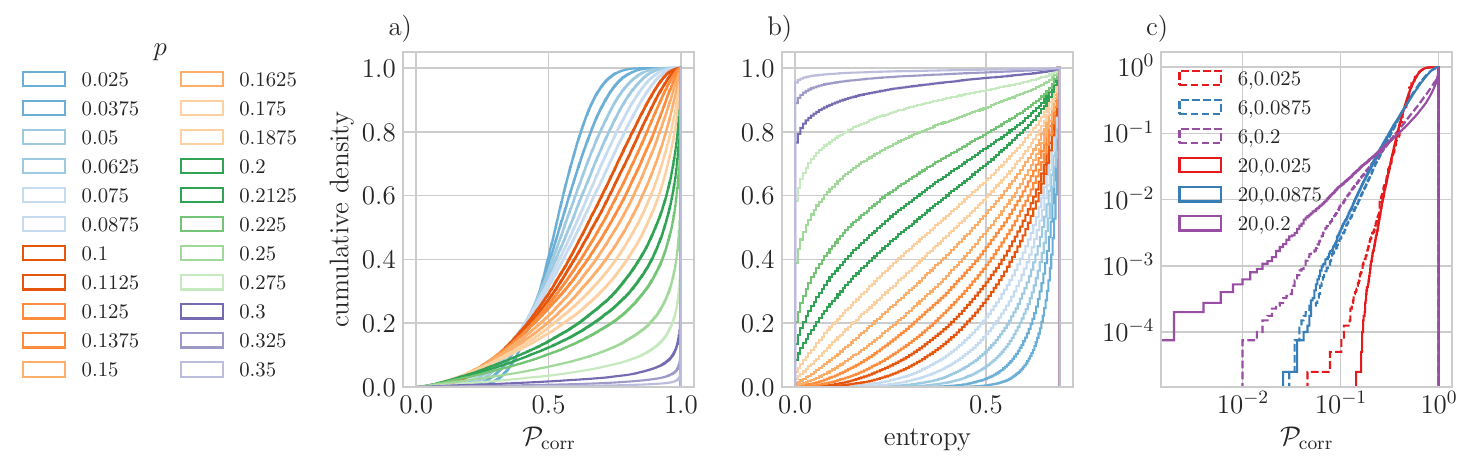}
    \caption{a) Cumulative histograms of ${\mathcal P}_{\rm corr}$ for varying $p$. System size is fixed at the largest studied $L=20$, and $40,000$ samples are drawn from the measurement distribution for each value of $p$. As discussed in the main text, the form of the distribution of ${\mathcal P}_{\rm corr}$ changes at $p_{\rm tails} \sim 0.1$ from a gaussian (${\rm erf}$ on a cumulative plot) to an exponential distribution. b) Cumulative distribution of the entropy of the classifier posterior distribution as a function of $p$. Distribution changes form at around $p_{\rm success}$, at which there is a transition in Eve's ability to determine the charge. c) Trends in the distribution of ${\mathcal P}_{\rm corr}$ and system size. Note the reversal of the trend in the tail around $p_{\rm tails}$.}
    \label{fig:add_data}
\end{figure}
\begin{figure}[t]
    \centering
    \includegraphics[width=0.7\linewidth]{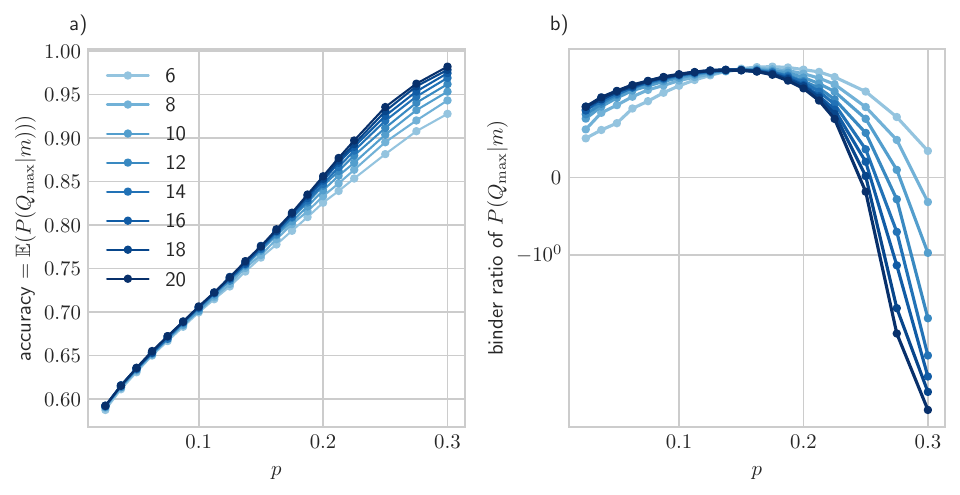}
    \caption{Transitions in the tails of the order parameter distribution. a) The (theoretical) accuracy, calculated from the same data, but as $\mathbb{E}(P(Q_{\rm max}|m))$, where $Q_{\rm max}(m) = \mathrm{argmax}_Q P(Q|m)$ is ``most likely label'' for a given measurement record, rather than as the fraction of correctly classified instances. The two should agree in the infinite data limit, since $P(Q_{\rm max} | m )$ is the correct classification rate for each measurement recrod. b) The binder ratio for this quantity (whose mean is the accuracy). The binder ratio exhibits a transition at $p \sim 0.15$, a point which coincides with neither of the transitions discussed in the main paper.}\label{fig:q_max}
\end{figure}
\section{Biased decoders}
How much better can we do with knowledge of the unitaries? 
Recall that the dataset is drawn from the joint distribution:
\begin{equation}
    p(m, y, u) = p(m | y, u) p(y) p(u) =  ||C(m, u)|Q_y\rangle ||^2 p(y) p(u).\label{eq:full_dist}
\end{equation}
Instead of considering $m, y$ pairs, consider instead pairs $(m, u), y$, that is, the task is now to predict a label given a tuple $m, u$ consisting of both measurement outcomes and unitary gate settings. 

There is a natural, optimal scheme given exponential resources. The probability distribution $p(y | m, u)$ follows from Eq.~\ref{eq:full_dist} by standard probability manipulations. Eq.~\ref{eq:full_dist} can be simulated exactly on a classical computer given knowledge of the unitaries and the measurement outcomes, with a cost that scales exponentially in the size of the data. 

More interestingly, by marginalising over fewer parameters from the Haar distribution of unitaries, we are left with a classifier that can include some information about each given unitary realisation. In performing this marginalisation, we can attain a classically efficient algorithm for decoding the measurement records. 
Importantly, this process of marginalisation yields a classifier that is optimal given only the remaining information. 

In the main text we consider the case where we know only the hopping probabilities from the unitary distribution. 
Consider the following parametrisation of the $U(1)$ symmetric two qubit unitaries: 
\begin{equation}
    U(\alpha, \rho, \psi, \chi) = \begin{pmatrix}
        1 & 0 & 0& 0 \\ 
        0 & e^{i(\alpha+\psi)}\sqrt{1-\xi} & e^{i(\alpha+\chi)} \sqrt{\xi} & 0 \\ 
        0 & -e^{i(\alpha-\chi)}\sqrt{\xi} & e^{i(\alpha-\psi)} \sqrt{1-\xi} & 0 \\
        0 & 0 & 0 & e^{{i\rho}}
    \end{pmatrix}.
\end{equation}
The Haar distribution we consider comes from drawing $\rho, \psi, \chi, \alpha \sim U(0, 2\pi)$ and $\xi \sim U(0, 1)$.
If in the average of Eq.~\ref{eq:proj} we average only over $\rho, \psi, \chi, \alpha$, the resulting transfer matrix
\begin{equation}
    {T}_{jk} = \bra{j} \otimes \bra{j} \mathcal{T} \ket{k} \otimes \ket{k} = \begin{pmatrix} 1 & 0 & 0 & 0 \\ 0 & 1-\xi & \xi & 0 \\ 0 & \xi & 1-\xi & 0 \\ 0 & 0 & 0 & 1 \end{pmatrix}_{jk},
\end{equation}
is precisely that of the biased decoder we discuss in the main paper. In Fig.~3 of the main paper (and Fig.~\ref{fig:antibiased} of the supplementary) we show that including information about the hopping probabilities suffices to shift the success transition downwards, allowing Eve to detect the charge at a lower critical measurement rate. 

Whether an efficient, optimal scheme exists that incorporates all of the information from the unitary gates remains an open question. 
Doing so would require efficiently evaluating the full distribution of Eq.~\ref{eq:full_dist}.
\subsection{Numerics}
We compare the performance of the biased decoder with two alternatives: the unbiased decoder, and the \emph{antibiased} decoder -- that with all of the hopping probabilities $\xi$ flipped $\rightarrow 1-\xi$. 
If including the effects of the biases improves the accuracy and confidence of the decoder, it should outperform both. This is in fact what we see (Fig.~\ref{fig:antibiased}). 
When compared to the unbiased decoder, the biased decoder has the same accuracy threshold, but a smaller success threshold. 
When comparing to the antibiased decoder, the biased decoder has the same \emph{success} threshold, but a smaller accuracy threshold. 
The biased decoder strikes a better balance between accuracy and confidence than either the unbiased or the antibiased decoder. 

\section{Percolation transition}
In this section we discuss the details of the percolation transition that occurs at $p\sim0.3$.

Consider each unitary in the circuit as a vertex in a diagonal lattice. On each edge of that lattice there are three measurement probabilities -- either the site is unmeasured, or the site is measured, in which case there are two possible outcomes: $0$, indicating the absence of a charge, and $1$, indicating the presence of a charge. The general layout is shown in Fig.~\ref{fig:dual}.

An important factor is that the conservation law implies that each measurement gives more information about the charge than just the value at a particular site. 
In particular, draw the dual lattice as indicated in gray dashed lines in Fig~\ref{fig:dual}. 
Then any simple loop that we can draw on this lattice must enclose exactly zero charge (since the charge entering the loop via one boundary must equal the charge leaving via the other. 
The same is not true for loops that enclose the whole system (for open boundary conditions, which we consider here, these amount to lines of measurements that touch both boundaries at any point). 
These loops must enclose the total charge of the initial state. 
By counting the measurement outcomes along this spanning line, we can determine the charge of the initial state. 
The existence of a spanning line of measurements is equivalent to the existence of a percolating cluster of measurements on the dual lattice. 

Importantly, this percolation problem is modified by the additional information given in each measurement.
By drawing loops that enclose a single unmeasured site, the value of that unmeasured site is completely constrained by charge conservation. 
We term the sites whose values we implicitly measure 'unmeasured sharp sites'. 
By adding unmeasured sharp sites to measured sites, the effective measurement rate is increased, and the percolation transition shifts to a lower value of $p$. 
Examples of these loops (considered as vertices on the primal lattice) are shown in Fig.~\ref{fig:dual}b).

Determining the charge from a percolating cluster is implicitly solved by our statistical mechanics model - since, if there is such a cluster, only one charge label will be consistent with the measurement outcomes. 
This percolation behaviour is visible in the histograms of ${\mathcal P}_{\rm corr}$ in the main paper (and in Fig.~\ref{fig:add_data}) as a shift to a binary distribution as the measurement rate crosses the percolation transition at $p \sim 0.3$.

\section{Variations}\label{sec:variations}
In this section we discuss two variations of the model described in the main text. 
\subsection{Product initial states}
The presentation in the main text focuses on discriminating the charge of circuits applied to balanced superpositions of charge states $\ket{Q}  = \frac{1}{\binom{N}{Q}} \sum_{i: Q(i) = Q} \ket{i}$.
These states are maximally entangled. 
The same effects described in the main text can be demonstrated starting from product initial states (Fig.~\ref{fig:neel}).

\subsection{Distinguishing many charge states}
The general problem we set up is that of distinguishing two charge states, $\ket{Q_0}$ and $\ket{Q_1}$. 
One can also consider a more general formulation, where the system starts off in a superposition of all charge states $\otimes_i^N \ket{+}_N$.
If we add a single terminal measurement of the global charge to the system, we can reframe the job of the eavesdropper as determining the charge that this measurement will collapse the system to.
The result is an $L$ class classification problem, wherein the job of the eavesdropper is to pick from the $L$ possible values of the charge. The decoder in this instance runs in the same way as described in the main paper for the original problem formulation, except that Eve now picks the most probable value from all $L$ charge possibilities. 
Fig.~\ref{fig:l_class} shows that this problem exhibits the same transitions as a function of measurement rate. 

\section{Additional Data}
\subsection{Cumulative distributions}
In Fig.~\ref{fig:add_data}a, b we present additional data on the transitions at $p_{\rm tails}$ and $p_{\rm success}$, showing the form of the cumulative distributions of ${\mathcal P}_{\rm corr}$ and the posterior distribution of the entropy of the classifier, showing transitions in behaviour at $p_{\rm tails}$ and $p_{\rm success}$ respectively. 

\subsection{The transition at $p_{\rm tails}$}
${\mathcal P}_{\rm corr}$, the quantity analysed in the main text, has a transition in the form of its distribution at some critical value $p_{\rm tails}$, which, by analysing the binder ratio (which is, up to a scaling and shift, the kurtosis of the order parameter distribution,~i.e. the weight in its tails), we located at $p_{\rm tails} \sim 0.1$.
The transition in the distribution ${\mathcal P}_{\rm corr}$ is not a transition in the averaged order parameter of the standard kind.
In fact, the average of ${\mathcal P}_{\rm corr}$, $\mathbb{E}({\mathcal P}_{\rm corr})$ can be shown to correspond to the exponential of the 2nd R\'enyi entropy of the posterior distribution of the classifier as follows.
Consider a fixed measurement record $m$, arising in the data with probability $p(m)$.
That value of $m$ might have been generated from label $Q_1$ (with probability $P(Q_1 | m)$), or it might have been generated from label $Q_0$ (with probability $1-P(Q_1 | m) = P(Q_0 | m)$. 
${\mathcal P}_{\rm corr}$ as defined in the paper corresponds to the probability of the label that the measurement record was actually generated from. 
So,
\begin{equation}
    \mathbb{E}({\mathcal P}_{\rm corr}) = \mathbb{E}_m \left[\mathbb{E}_{q | m} ( {\mathcal P}_{\rm corr} )\right] = \mathbb{E}_m \left[ P(Q_0 | m)^2 + P(Q_1 | m)^2 \right]] = \mathbb{E}_m \left[ e^{-2 \mathcal{R}_2(m)} \right]]
\end{equation}
where $\mathcal{R}_2(m)$ is the 2nd renyi entropy of the charge distribution the classifier for the measurement record $m$. 
This quantity is actually an order parameter for the success transition at $p_{\rm success}$, being related to the Shannon entropy of the charge distribution. 
We therefore interpret the transition at $p_{\rm tails}$ is a discontinuous change in the behaviour of the total distribution of ${\mathcal P}_{\rm corr}$ (in the thermodynamic limit), that does not show up in the average. 

Fig.~\ref{fig:add_data}c presents data supporting the validity of this interpretation, demonstrating how the behaviour of the distribution of ${\mathcal P}_{\rm corr}$ changes with system size. 
A clear flow reversal is demonstrated at $p_{\rm tails}$.

Fig.~\ref{fig:q_max} gives further perspective on the transition at $p_{\rm tails}$. 
Instead of considering ${\mathcal P}_{\rm corr}$, the probability associated with the ground truth label, we can look instead at the distribution of $P(Q_{\rm max}|m)$, the probability associated with the 'most likely' label - that is, the probability the model assigns to its label prediction. 
Because the classifier is optimal, this quantity averages to the accuracy - since the probability the model assigns to its predicted labels matches the distribution of those labels in the dataset, the probability of the ``most likely'' label corresponds to (one minus) the misclassification rate (the rate at which ``unlikely'' measurement records show up in the dataset).
We can see in Fig.~\ref{fig:q_max} that this theoretical accuracy indeed tracks the accuracy calculated empirically in Fig.~2 of the main text.
The quantity $P(Q_{\rm max}|m)$ also exhibits a transition in the form of its distribution. 
This is shown in Fig.~\ref{fig:q_max}b).

\section{Numerical Details}
Measurement records were generated from an exact simulation of the dynamics of the quantum state vector for time $L$. All of the data presented is for $20,000$ samples for each charge, measurement rate, size combination. Open boundary conditions were used for both data generation and decoding. 

The decoder performs a TEBD simulation of Eq.~3, (main paper) using the tooling provided by the quimb \cite{gray2018quimb} library. The TEBD threshold was set to $1\times 10^{-10}$.
\bibliography{decoding}